\documentstyle{mn}

\input epsf

\begin{document}

\def\Lya{Ly$\alpha\ $}
\def\Lyb{Ly$\beta\ $}
\def\Lyg{Ly$\gamma\ $}
\def\Lyd{Ly$\delta\ $}
\def\Lye{Ly$\epsilon\ $}
\def\LCDM{$\Lambda$CDM\ }
\def\HI{\hbox{H~$\rm \scriptstyle I\ $}}
\def\HII{\hbox{H~$\rm \scriptstyle II\ $}}
\def\DI{\hbox{D~$\rm \scriptstyle I\ $}}
\def\HeI{\hbox{He~$\rm \scriptstyle I\ $}}
\def\HeII{\hbox{He~$\rm \scriptstyle II\ $}}
\def\HeIII{\hbox{He~$\rm \scriptstyle III\ $}}
\def\CIV{\hbox{C~$\rm \scriptstyle IV\ $}}
\def\NHI{N_{\rm HI}}
\def\NHeII{N_{\rm HeII}}
\def\cm2{\,{\rm cm$^{-2}$}\,}
\def\cm3{\,{\rm cm$^{-3}$}\,}
\def\kms{\,{\rm km\,s$^{-1}$}\,}
\def\skm{\,({\rm km\,s$^{-1}$})$^{-1}$\,}
\def\kmsmpc{\,{\rm km\,s$^{-1}$\,Mpc$^{-1}$}\,}
\def\hmpc{\,h^{-1}{\rm \,Mpc}\,}
\def\mpch{\,h{\rm \,Mpc}^{-1}\,}
\def\hkpc{\,h^{-1}{\rm \,kpc}\,}
\def\ev{\,{\rm eV\ }}
\def\kel{\,{\rm K\ }}
\def\intunits{\,{\rm ergs\,s^{-1}\,cm^{-2}\,Hz^{-1}\,sr^{-1}}}
\def\emunits{\,h\,{\rm erg\,s^{-1}\,Hz^{-1}\,Mpc^{-3}}}
\def\ltsima{$\; \buildrel < \over \sim \;$}
\def\lsim{\lower.5ex\hbox{\ltsima}}
\def\gtsima{$\; \buildrel > \over \sim \;$}
\def\gsim{\lower.5ex\hbox{\gtsima}}
\def\etal{{ et~al.~}}
\def\aj{AJ}
\def\apj{ApJ}
\def\apjs{ApJS}
\def\apsss{Ap\&SSS}
\def\araa{ARA\&A}
\def\mnras{MNRAS}

\journal{Preprint-03}

\title{Constraints on the UV metagalactic emissivity using
the Ly$\alpha$ forest}

\author[A. Meiksin and M. White]{Avery Meiksin${}^{1}$, Martin White${}^{2}$ \\
${}^1$Institute for Astronomy, University of Edinburgh,
Blackford Hill, Edinburgh\ EH9\ 3HJ, UK \\
${}^2$Departments of Astronomy and Physics, University of California,
Berkeley, CA 94720, USA}

\pubyear{2003}

\maketitle

\begin{abstract}
Numerical hydrodynamical simulations have proven a successful means of
reproducing many of the statistical properties of the \Lya forest as
measured in high redshift quasar spectra. The source of ionization of
the Intergalactic Medium (IGM), however, remains unknown. We
investigate how the \Lya forest may be used to probe the nature of the
sources. We show that the attenuation of Lyman continuum photons by
the IGM depends sensitively on the emissivity of the sources,
permitting a strong constraint to be set on the required emissivity to
match the measured values of the mean IGM \Lya optical depth. We find
that, within the observational errors, QSO sources alone are able to
account for the required UV background at $z\gsim4$. By contrast, the
emissivity of Lyman Break Galaxies must decline sharply with redshift,
compared with the estimated emissivity at $z\approx3$, so as not to
over-produce the UV background and drive the mean \Lya optical depth
to too low values. We also investigate the effect of fluctuations in
the UV background, as would arise if QSOs dominated.  To this end, we
derive the distribution function of the background radiation field
produced by discrete sources in an infinite universe, including the
effects of attenuation by an intervening absorbing medium. We show
that for $z\gsim5$, the fluctuations significantly boost the mean \Lya
optical depth, and so increase the estimate for the mean ionization
rate required to match the measured mean \Lya optical depths. The
fluctuations will also result in large spatial correlations in the
ionization level of the IGM. We show that the large mean \Lya optical
depth measured at $z\approx6$ suggests such large correlations will be
present if QSOs dominate the UV background. A secondary, smaller
effect of the UV background fluctuations is a distortion of the pixel
flux distribution. While the effect on the distribution may be too
small to detect with existing telescopes, it may be measurable with
the extremely large telescopes planned for the future. We also show
that if QSOs dominate the UV background at $z\approx6$, then they will
be sufficient in number to rejuvenate the ionization of a previously
ionized IGM if it has not yet fully recombined.
\end{abstract}

%%%%%%%%%%%%%%%%%%%%%%%%%%%%%%%%%%%%%%%%%%%%%%%%%%%%%%%%%%%%%%%%%%%%%%%%%%%%%%%
\section{Introduction}
\label{sec:introduction}

Numerical simulations of structure formation in the universe incorporating
hydrodynamics in Cold Dark Matter (CDM) dominated cosmologies have proven
very successful in reproducing the statistical properties of the \Lya forest
as measured in high redshift Quasi-Stellar Object (QSO) spectra
(Cen \etal 1994; Zhang, Anninos \& Norman 1995; Hernquist \etal 1996;
Zhang \etal 1997; Bond \& Wadsley 1997; Theuns, Leonard \& Efstathiou 1998a).
The high level of statistical agreement, as high as a few percent in the
cumulative spectral flux distributions (Meiksin \etal 2001), suggests that
the models are capturing the essential physical nature of these highly ionized
absorbing structures.

Still unknown is the nature of the sources of photoionization. The most
likely candidates are photoionization by Quasi-Stellar Objects (QSOs) and
stars in young forming galaxies or globular clusters. Estimates of the UV
background required to reproduce the measured \Lya optical depths of the
\Lya forest while staying within the nucleosynthesis bounds of the cosmic
baryon density agree within a factor of a few with estimates of the QSO
contribution to the UV background (Meiksin \etal 2001). This suggests that
QSO sources contribute significantly to the UV background, and may even
dominate over all other sources. A contribution will also arise from
star forming galaxies at $z>3$, from which Lyman-continuum photons have
been detected (Steidel, Pettini \& Adelberger 2001). The total contribution
from QSOs, or Active Galactic Nuclei (AGN) more generally, and star-forming
galaxies, however, remains uncertain because of the flux limitations of the
surveys that detect them at these high redshifts. Generally only the brightest
sources enter the surveys, leaving the possibility of a large contribution that
has gone undetected. Distinguishing the relative contributions of young stars
vs AGN is thus not possible based soley on the average detected contributions
of each to the UV metagalactic background.

The discreteness of the sources presents an opportunity for distinguishing
between the possible natures of the dominant sources. If star forming
galaxies or globular star clusters contribute a comparable amount or
more of ionizing photons to the metagalactic background as do QSO
sources, then they must have a much higher spatial density than QSOs
because of their relatively weaker individual luminosities. The
resulting UV background would then be fairly homogeneous, except for
local modulations arising from the large scale structure of their
spatial distribution. By contrast, large fluctuations in the UV
background are expected at high redshifts ($z\gsim4$) if QSO sources
dominate the background because of the rapidly growing optical depth
of the Intergalactic Medium (IGM) to Lyman continuum photons and the
rapidly diminishing number density of QSO sources (Zuo 1992a). The
fluctuations in the QSO contribution will become particularly
pronounced at $z>5$, where the mean \Lya optical depth exceeds unity
(Fan \etal 2001b; Becker \etal 2001), as we show here.

The role of fluctuations in the UV background on the absorption
properties of the \Lya forest have been largely unexplored (but see
Croft \etal 2002).  In this paper, we investigate the effect of the
fluctuations on the 1-pt flux distribution of the \Lya forest. We
examine the effect of the background fluctuations on the power
spectrum of the \Lya forest in a companion paper (Meiksin \& White
2003). We base the contributions of the QSOs to the UV background and
its fluctuations on the QSO luminosity functions obtained by the 2dF
(Boyle \etal 2000) and the SDSS (Fan \etal 2001a, 2001b, 2003).

Because of the large computational overhead in incorporating
proper hydrodynamics into an N-body code, we perform simulations of
the \Lya forest using a pure gravitational Particle-Mesh N-body code
(PM simulations). These simulations have been shown to agree well with
the full hydrodynamics computations (Meiksin \& White 2001), and have
the advantages of permitting both larger simulations to be run as well as
greater numbers of them. Ultimately, a detailed
comparison with the data will require solving the full set of hydrodynamics
equations.

This paper is organized as follows:\ In section~\ref{sec:sims}, we discuss
our simulations of the \Lya forest and model assumptions; estimates of the
metagalactic emissivity are given in section~\ref{sec:emissivity}; we derive
the distribution function for the UV background fluctuations in
section~\ref{sec:flucs}; in section~\ref{sec:flux} we discuss the pixel flux
distribution; we finally summarise our results in section~\ref{sec:summary}.

%%%%%%%%%%%%%%%%%%%%%%%%%%%%%%%%%%%%%%%%%%%%%%%%%%%%%%%%%%%%%%%%%%%%%%%%%%%%%%%
\section{The simulations and models}
\label{sec:sims}

\subsection{Particle-mesh simulations}
\label{sec:PM}

We generate spectra of the \Lya forest from pure DM simulations,
mimicking the temperature of the gas using a polytropic equation of
state. Comparisons with full hydrodynamics simulations show that
simulations recover the principal spectral properties of the \Lya
forest to an accuracy of 10\% or better in the cumulative
distributions of flux, wavelet coefficients, \HI column density, and
Doppler parameter (Meiksin \& White 2001).

The evolution of the (dark) matter is computed using a parallel
implementation of a PM code. The computational volume is chosen to be
periodic with side length 1, and we take as our time coordinate the
log of the scale factor, $\log a$, where $a\equiv (1+z)^{-1}$.
Velocities are measured in units of the expansion velocity across the
box $aHL_{\rm box}$ of comoving side $L_{\rm box}$, where $H$ is the
Hubble parameter fixed by the Friedmann equation
\begin{equation}
  H^2 \equiv H_0^2\left( {\Omega_M\over a^3} + \Omega_\Lambda
  + {\Omega_K\over a^2} \right)
\end{equation}
in terms of the Hubble constant $H_0$ and the densities in matter
($\Omega_M$), cosmological constant ($\Omega_\Lambda$) and curvature
($\Omega_K$), in units of the critical density. The density is
defined from the particle positions by assigning particles to a
regular cartesian grid using the Cloud-In-Cell (CIC) scheme (Hockney
\& Eastwood~\cite{HocEas}).  The Fourier Transform is taken and the
force computed using the kernel $\vec{k}/k^2$.  We use a second order
leap frog method to integrate the equations.  The relevant positions
are predicted at a half time step and used to calculate the
accelerations which modify the velocities.  The time step is
dynamically chosen as a small fraction of the cell crossing time with
a maximum size of $\Delta\log a=2\%$.

The initial conditions were created by displacing the particles from a uniform
grid using the Zel'dovich approximation.  The initial conditions were Gaussian,
however to avoid large fluctuations between different realizations we only
accepted those whose low-$k$ modes closely tracked the ``mean''.

Given a set of final particle positions and velocities, we compute the spectra
as follows.  First the density and density-weighted line-of-sight velocity
are computed on a grid (using CIC interpolation as above) and Gaussian smoothed
using FFT techniques in order to adequately sample the velocity field.
This forms the fundamental data set.
A regular grid of $32\times 32$ sightlines is drawn through the box,
parallel to the box sides.  Along each sightline we integrate (in real space)
to find $\tau(u)$ at a given velocity $u$.  Specifically we define
\begin{equation}
  \tau(u) = \int dx A(x) \left[ {\rho(x)\over\bar{\rho}} \right]^2 T(x)^{-0.7}
            b^{-1} e^{-(u-u_0)^2/b^2}
\label{eq:taudef}
\end{equation}
where $u_0=xaHL_{\rm box}+v_{\rm los}$ and $b=\sqrt{2k_B T/ m_{\rm
H}}$ is the Doppler parameter, where $m_{\rm H}$ is the mass of a
hydrogen atom.  The flux at velocity $u$ is $\exp(-\tau)$. The
integration variable $x$ indicates the distance along the box in terms
of the expansion velocity across the box. In terms of the baryon
density parameter $\Omega_b$, hydrogen baryonic mass fraction $X$, the
metagalactic photoionization rate $\Gamma_{-12}$, in units of
$10^{-12}\, {\rm photons\, s^{-1}}$, and the number $f_e$ of electrons per
hydrogen nucleus, the coefficient $A$ is given by:
\begin{equation}
A(x)=18.0f_e\left(\frac{X}{0.76}\right)^2
\left(\frac{\Omega_bh^2}{0.02}\right)^2\Gamma_{-12}^{-1}(x)L_{\rm box}(1+z)^5,
\label{eq:Adef}
\end{equation}
where $L_{\rm box}$ is in Mpc and $b$ in eq.~(\ref{eq:taudef}) is in
\kms. The explicit $x$ dependence of $A$ accounts for any spatial
fluctuations in the background ionization rate $\Gamma_{-12}(x)$.
The electron fraction is given by $f_e=1+(1-X)/(2X)\approx1.16$ ($X=0.76$)
for fully ionized helium, and $f_e=1+(1-X)/(4X)\approx1.08$ if the helium
is only singly ionized, a possibility at the high redshifts we consider.
We shall adopt $f_e=1.1$ for simplicity, noting that a different value for
$f_e$ is effectively absorbed into the ionization rate.

In the absence of UV background fluctuations, $A$ is constant and
is iteratively adjusted to obtain a predefined
$\bar\tau_\alpha\equiv-\log\left\langle\exp(-\tau)\right\rangle$.  To
account for a fluctuating UV background field, we replace $A$ by
$A_{-12}/ [\Gamma_{\rm S, -12} + \bar\Gamma_{\rm QSO,-12}
(1+\delta_{\rm QSO})]$, where $\bar\Gamma_{\rm QSO, -12}$ is the
average photoionization rate produced by the QSO sources, in units of
$10^{-12}\, {\rm s^{-1}}$, $A_{-12}$ is the value of $A$ for $\Gamma_{-12}=1$,
and $\delta_{\rm QSO}$ accounts for the spatial fluctuations in the QSO
contribution. The values for $\delta_{\rm QSO}$ are drawn randomly from
the distribution functions for the background intensity derived below.
We shall consider two cases
for the sources of the UV background. In the first, we assume the
only sources are QSOs, so that $\Gamma_{\rm S}=0$. In the second,
we shall allow for a smooth component $\Gamma_{\rm S}$
to the UV background. This component will partly be present due to
the re-radiation of ionizing photons through recombination in the \Lya forest
itself, at a level by as much as $(\alpha_A-\alpha_B)/\alpha_A\approx40\%$
of the QSO contribution (Meiksin \& Madau 1993; Haardt \& Madau 1996).
Because of structure in the \Lya forest, the
re-radiated component will also fluctuate. We neglect this contribution
to the fluctuations since they likely will be less than the direct contribution
from QSOs. Other possible contributions to the ionizing background include
galaxies, like Lyman Break Galaxies, and smaller stellar systems like
intergalactic star clusters. In this case, the value of $\Gamma_{\rm S}$
is iterated until agreement with $\bar\tau_\alpha$ is found.

In evaluating Eq.~(\ref{eq:taudef}) we assume a power-law relation between
the gas density and temperature
\begin{equation}
  T \equiv T_0 \left( {\rho\over\bar{\rho}} \right)^{\gamma-1}.
\label{eq:eos}
\end{equation}
In practice, $T_0$ and $\gamma$ will depend on the reionization
history of the IGM, but we shall fix $T_0=2\times10^4\kel$ and $\gamma=1.5$
throughout for most of our models. The value we use for $T_0$ is larger than
used by Meiksin \& White (2001),
to account for the broad lines measured over the full range of line-centre
optical depths near $z=3.5$ (Meiksin \etal 2001). An effort to calibrate
the evolution of $T_0$ and $\gamma$ was undertaken by Schaye \etal (2000).
The origin of the extra broadening is unknown. 
A plausible explanation is late \HeII reionization, but it may be due to
turbulence driven by galactic (or star cluster) winds, in which case it
may well be present to redshifts as high as 6 or even higher. Since we
are not concerned with detailed matching to the absorption statistics
of particular spectra in this
paper, we have not included a complicated evolutionary scenario for
$T_0$ and $\gamma$. Instead we simply consider a spread in $T_0$ by
also considering cases with $T_0=10^4\kel$, noting that the equation of
state is only an approximate description of the IGM temperature and is in any
case valid only for overdense regions (Meiksin 1994; Theuns \etal 1998b).

\subsection{Cosmological model}
\label{sec:models}

We consider a flat cosmological model with
$\Omega_M=0.30$, $\Omega_\Lambda=0.70$,
$\Omega_b=0.041$, $h=H_0/ 100$\kmsmpc~$=0.70$,
and slope of the primordial density
perturbation power spectrum $n=1.05$. The model is COBE normalized, so that
the fluctuation normalization at $z=0$ in a sphere of radius $8h^{-1}$ Mpc
is $\sigma_{8h^{-1}}=0.970$, and the fluctuation normalization at $z=3$
spherically filtered on the scale of the Jeans length for $T=2\times10^4\kel$
gas is $\sigma_J(z=3)=1.59$. The model is consistent with current constraints
from the CMB (Stompor \etal 2001; Netterfield \etal 2002; Pryke \etal 2002),
Big Bang Nucleosynthesis (O'Meara \etal 2001), and large-scale structure
(Percival \etal 2001; Szalay \etal 2001), as well as the constraint on the size
of mass fluctuations on the scale of the Jeans length at $2<z<4$ imposed
by the measured flux per pixel distribution of the \Lya forest
(Meiksin \etal 2001).
We use $512^3$ particles on a $1024^3$ mesh in a box of comoving side
$L_{\rm box}=25 h^{-1}\, {\rm Mpc}$.

\subsection{Sources of the UV background}
\label{sec:sources}

We consider two sources of the UV background: QSOs and stellar, such as from
Lyman Break Galaxies (LBGs).

We parametrize the QSO comoving luminosity function using the form introduced
by Boyle \etal (1988),
\begin{equation}
\Phi(M,z)=\frac{\Phi^*}{10^{0.4(\beta_1-1)[M^*(z)-M]} +
10^{0.4(\beta_2-1)[M^*(z)-M]}},
\label{eq:BLF}
\end{equation}
where $M^*(z)$ represents an evolving break magnitude. The results of
the 2dF QSO survey show that the QSO counts for $z<2.3$ can be
acceptably fit by a a pure luminosity evolution model in which
$\Phi^*$ is independent of redshift, while the break magnitude
evolves according to $M^*(z)=M^*(0) - 2.5(k_1 z + k_2 z^2)$. For
$\Omega_M=0.3$, $\Omega_\Lambda=0.7$, and $h=0.5$, Boyle \etal (2000)
obtain for the $B-$band luminosity function
$\beta_1=3.41$, $\beta_2=1.58$, $\Phi^*=0.36\times10^{-6}\,
{\rm Mpc^{-3}\, mag^{-1}}$, $M_B^*(0)=-22.65$, $k_1=1.36$, and $k_2=-0.27$.

At $z>2.3$, the QSO luminosity function is less well determined. We extend
the luminosity function to higher redshifts using the results of the Sloan
Digital Sky Survey for $z>3.5$ (Fan \etal 2001a, 2001b, 2003). The survey
limit is too bright to probe to fainter levels than the break, but shows
a shallowing in the bright end of the luminosity function to $\beta_1=2.58$, so
that the pure luminosity evolution model may no longer apply at the bright end.
Since the pure luminosity model with $\beta_1=3.4$ is ruled out at only the
2-2.5~$\sigma$ level, and possibly only for the brightest QSOs ($M_{1450}<-27$)
(Fan et al. 2001a, 2001b, 2003), we shall consider the pure luminosity
evolution case as well. Since for the values of $\beta_1$ and $\beta_2$ we
shall consider, eq.~(\ref{eq:BLF}) predicts that the UV background is dominated
by sources with absolute magnitudes near or fainter than $M^*$, the
contribution of QSOs is poorly constrained by observations. We conservatively
assume that the faint end of the luminosity function continues to follow pure
luminosity evolution with $\beta_2=1.58$, and fix the break luminosity $M^*(z)$
in eq.~(\ref{eq:BLF}) to match the bright counts measured by the Sloan.

\begin{figure}
\begin{center}
\leavevmode \epsfxsize=3.3in \epsfbox{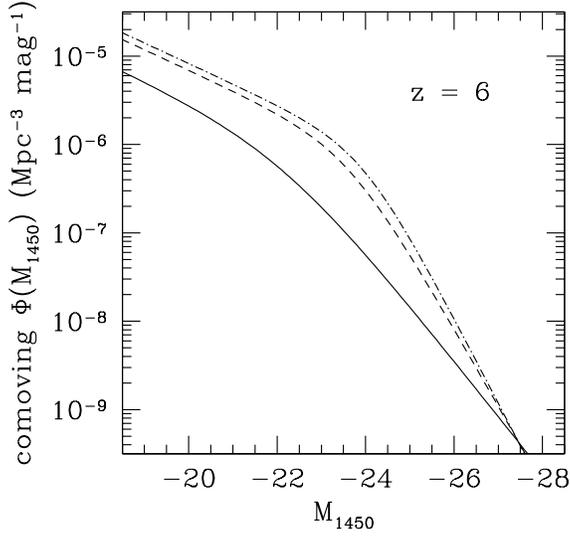}
\end{center}
\caption{The comoving QSO luminosity function at $z=6$ for the $\Lambda$CDM
model with a bright end slope of $\beta_1=2.58$ (solid), 3.20 (dashed),
and 3.41 (dot-dashed), with the break magnitude $M^*_{1450}$ adjusted to
match the QSO counts at the bright end measured by the SDSS.}
\label{fig:qsolf}
\end{figure}

The luminosity function reported by Fan \etal (2001a, 2001b, 2002b) is in
terms of the absolute magnitude $M_{1450}$ at $1450$~\AA. To estimate the
contribution of the QSOs to the UV background, we need to convert
these values to luminosities at the Lyman edge. We adopt the spectral
shape of radio quiet QSOs as determined at lower redshift by
measurements using the {\it HST} FOS (Zheng \etal 1997). For
$\lambda<1050$~\AA, we take $f_\nu\propto \nu^{-1.8}$, while for
$1050$~\AA$<\lambda<1450$~\AA, we take $f_\nu\propto \nu^{-0.99}$. The
resulting break values $M^*_L$ at the Lyman edge for $\beta_1=2.58$ and
3.41 are given in Table~\ref{tab:LF}. (The averages over the luminosity
function have been performed over the range $M_{1450}<-18.5$, and the
results adjusted to $h=0.7$ for our $\Lambda$CDM model.)
For $\beta_1=2.58$, we find the proper emissivity decreases
with redshift for $z\ge4$ approximately as $(1+z)^{-1}$. We also give values
for the proper emissivity for $\beta_1=3.2$. This is less than 2~$\sigma$
away from the fits of Fan \etal (2001a) and is able to match the emissivity
required to reproduce the measured values of $\bar\tau_\alpha$ for
$4<z<6$ using QSO sources alone. For this case, the proper emissivity evolves
approximately as $(1+z)^{1/4}$ for $z\ge4$. We show the resulting comoving QSO
luminosity functions at $z=6$ for $\beta_1=2.58$, 3.20 and 3.41.

We also consider a possible contribution from Lyman Break Galaxies or other
stellar sources of Lyman continuum photons. Because
UV continuum radiation shortward of the Lyman break is only observed from the
brightest LBGs (Steidel et al. 2001), the full contribution of LBGs to the
ionizing background is uncertain by a factor of a few for $z\approx3$, and
even more uncertain at higher redshifts. The spectral index of the emission,
which affects the total ionization rate, is also highly uncertain, depending
on the processing of the continuum radiation through the interstellar medium
of the galaxy and the IGM, and on the stellar populations of the galaxy. For
the estimated ages of $10^{8\pm0.5}$~yr (Papovich, Dickinson \& Ferguson 2001),
the spectral shape is especially age-sensitive (Fioc \& Rocca-Volmerange 1997;
Bruzual 2001). For simplicity, we adopt a spectrally flat proper emissivity
near the Lyman edge of $\epsilon_\nu^{\rm LBG}\approx8\times10^{27}\emunits$
at $z\approx3$ (Steidel \etal 2001), noting that this estimate was for an
Einstein-deSitter universe, and so somewhat overestimates the contribution in
the $\Lambda$-dominated universes we consider here. The uncertainty in the
spectral index introduces an uncertainty in the photoionization rate of
$30-50\%$.

%%%%%%%%%%%%%%%%%%%%%%%%%%%%%%%%%%%%%%%%%%%%%%%%%%%%%%%%%%%%%%%%%%%%%%%%%%%%%%%
\section{The required metagalactic emissivity}
\label{sec:emissivity}

Within the context of any given set of cosmological parameters, the flux
distribution of the \Lya forest predicted by numerical simulations is
fully specified except for one uncertainty, the magnitude of the UV
background responsible for the photoionization. Generally this is fixed
by requiring the simulations to reproduce the measured average flux, or
equivalently, the mean \Lya optical depth $\bar\tau_\alpha.$
Usually this is done by specifying the required mean
photoionization rate $\Gamma$ per neutral hydrogen atom. In the presence of
fluctuations in the UV background, however, the full distribution
of $\Gamma$ should be used, since $\bar\tau_\alpha$ depends non-linearly
on $\Gamma$. This is particularly important when the distribution is
broad. We predict the distribution below and will take it into account
there. In this section, to simplify the calculation, we assume a constant
value of $\Gamma$ at any given redshift.

The photoionization rate is related to the emissivity of the sources through
\begin{equation}
\Gamma=\int_{\nu_L}^\infty d\nu\, \frac{4\pi J_\nu}{h\nu}\sigma_\nu,
\label{eq:Gamma}
\end{equation}
where $J_\nu$ is the angle-averaged intensity of the UV background at frequency
$\nu$, $\sigma_\nu\approx\sigma_L(\nu/\nu_L)^{-3}$, is the photoionization
cross-section, and $\nu_L$ is the frequency at the Lyman edge. The intensity
$J_\nu$ is given by the proper emissivity $\epsilon_\nu(z)$ of the sources
according to (eg, Meiksin \& Madau 1993),
\begin{equation}
J_\nu(z)=\frac{1}{4\pi}\int_z^{z_{\rm on}} dz'
\, \left(\frac{1+z}{1+z'}\right)^3
\epsilon_{\nu'}(z')\langle\exp\left[-\tau_\nu(z,z')\right]\rangle
\frac{dl_p}{dz'},
\label{eq:Jnu}
\end{equation}
where $\langle\exp\left[-\tau_\nu(z,z')\right]\rangle$ is the average
attenuation produced at $\nu$ due to intervening photoelectric absorption
by gas between redshifts $z$ and $z'$, $dl_p/dz'$ is the differential proper
cosmological line element, $\nu'=\nu(1+z')/(1+z)$, and it is assumed that the
sources have turned on at some finite redshift $z_{\rm on}$. At high redshift
in a flat universe, $dl_p/dz\approx(c/H_0)\Omega_M^{-1/2}(1+z)^{-5/2}$.

Following Zuo (1992b) and Zuo \& Phinney (1993), we introduce the attenuation
length $r_0$, defined by the approximation
$\langle\exp\left[-\tau_\nu(z,z')\right]\rangle \approx
\exp\left[-l_p(z,z')/r_0\right]$, where $l_p(z,z')$ is the proper distance
between redshifts $z$ and $z'$. Since our simulations do not include the
effects of radiative transfer, the full contribution of Lyman Limit Systems is
not accounted for. Their contribution to the total Lyman continuum optical
depth from the IGM, however, becomes small compared with that from the \Lya
forest itself at the high redshifts we consider here (Meiksin \& Madau 1993;
Madau \& Haardt 1996). As a test of their possible contribution, we replaced
the flux in pixels with $\tau_L>1$ by zero (fully blackening them), and found
this had a negligible effect on the values of $r_0$ obtained.

It is informative to consider two limiting cases for the contribution of the
sources to $J_\nu$:\ 1.\ absorption limited (AL), for which the attenuation
length is much smaller than the horizon,
and 2.\ cosmologically limited (CL), for which the attenuation length much
exceeds the horizon. For the AL case, the integral in eq.~(\ref{eq:Jnu})
reduces to $J_\nu=\epsilon_\nu(z)r_0/4\pi$. For the CL case, attenuation by the
IGM is negligible, and the integral is limited by cosmological expansion and
the age of the sources (avoiding Olber's paradox). In this case it is useful
to represent the redshift evolution and frequency dependence of the
emissivity as power laws:\ $\epsilon_\nu(z)=
\epsilon_L(1+z)^\eta(\nu/\nu_L)^{-\alpha_{\rm S}}$. Using the approximation
above for $dl_p/dz$, we then obtain $J_\nu(z) = \epsilon_\nu(z)l_{\rm U}/4\pi$,
where the effective proper distance in the universe over which sources
contribute to $J_\nu(z)$ is
\begin{equation}
l_{\rm U}=\frac{c}{H_0\Omega_M^{1/2}}(1+z)^{-3/2}
\frac{1-\left[(1+z)/(1+z_{\rm on})\right]^{9/2+\alpha_{\rm S}-\eta}}
{9/2+\alpha_{\rm S}-\eta}.
\label{eq:lU}
\end{equation}
Since the mean
\Lya optical depth constrains the photoionization rate $\Gamma$ through the
simulations rather than the emissivity directly, we re-express these results
in terms of $\Gamma$. While the above approximations express the
magnitude of the emissivity at the Lyman edge, they do not take into account
the effect of attenuation at shorter wavelengths in case AL. The effect of
intervening gas is to harden the spectrum, since high energy photons are able
to travel further than photons near the Lyman edge before they are
photoelectrically absorbed (Haardt \& Madau 1996). We account for the
hardening by introducing the metagalactic spectral index $\alpha_{\rm MG}$,
which replaces $\alpha_S$ for the AL case, but not for the CL case.
The two limiting cases may then be combined into the approximation
\begin{equation}
\Gamma(z)\approx\frac{\epsilon_L(z)\sigma_L}{h_P(3+\alpha_{\rm MG})}r_{\rm eff},
\label{eq:Gammap}
\end{equation}
where $h_P$ is the Planck constant, $r_{\rm eff}=1/(1/ r_0 + 1/ l_{\rm H})$,
and $l_{\rm H}=[(3+\alpha_{\rm MG})/(3+\alpha_S)]l_{\rm U}$. The cases AL and
CL then correspond to the respective limits $r_0\ll l_{\rm H}$ and
$r_0\gg l_{\rm H}$.

\begin{figure}
\begin{center}
\leavevmode \epsfxsize=3.3in \epsfbox{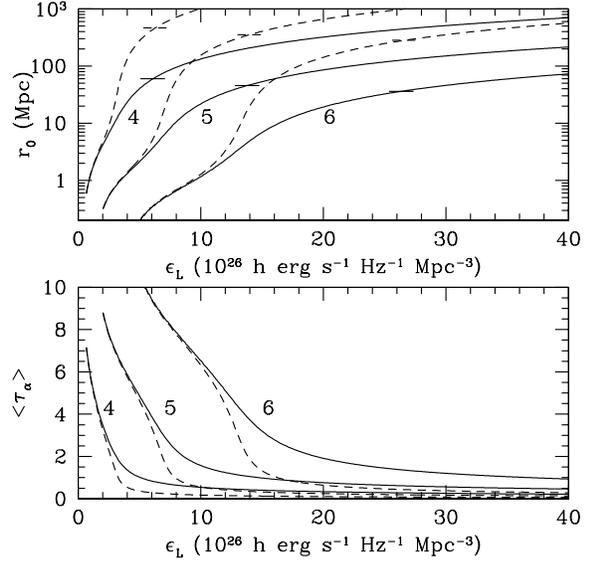}
\end{center}
\caption{The dependence of the attenuation length $r_0$ (top panel) and mean
\Lya optical depth $\bar\tau_\alpha$ (bottom panel) on the average emissivity
at the Lyman edge for the $\Lambda$CDM model at $z=4, 5$, and 6, as indicated.
It is assumed that the (proper) emissivity evolves according to
$\epsilon_\nu=\epsilon_L(1+z)^{-1}
(\nu/\nu_L)^{-1.8}$ for $z>4$, corresponding to QSOs (solid lines) or as
$\epsilon_\nu=\epsilon_L(1+z)^3$, corresponding to a constant comoving
source density (dashed lines). The horizontal bars in the upper panel mark
the effective luminosity horizons $l_{\rm H}$ of the sources (see text).}
\label{fig:emiss_r0_tau}
\end{figure}

\begin{figure}
\begin{center}
\leavevmode \epsfxsize=3.3in \epsfbox{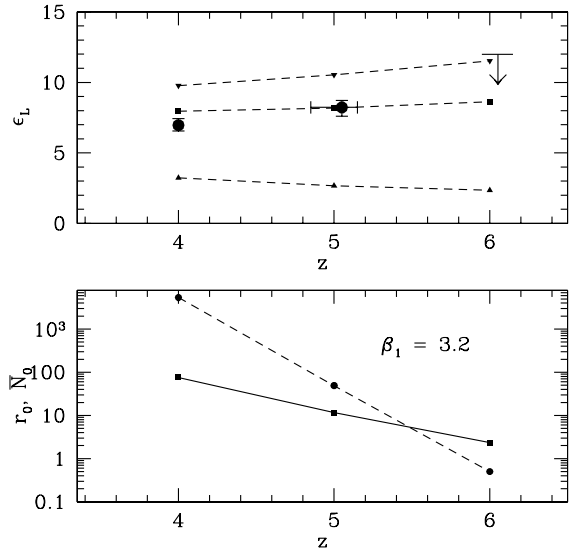}
\end{center}
\caption{(Top panel) The emissivity $\epsilon_L$ at the Lyman edge required to
match the measured mean \Lya optical depth $\bar\tau_\alpha$ at $z=4$, 5 and 6
(Becker \etal 2001; Songaila \& Cowie 2002). The units of the emissivity are
$10^{26}\emunits$. The estimates based on the $\Lambda$CDM model are shown
(points), with the error bars corresponding to the redshift ranges and
measurement errors on $\bar\tau_\alpha$. Only an upper limit is shown at $z=6$,
corresponding the the measured $1\sigma$ lower limit on
$\bar\tau_\alpha$. The estimates for the emissivity from QSO sources alone are
also shown, for $\beta_1=2.58$ (triangles), 3.20 (squares) and 3.41 (inverted
triangles). (Bottom panel) The attenuation length $r_0$, in proper Mpc (solid
line) and the mean number $\bar N_0$ of QSO sources within the attenuation
volume (dashed line) predicted by the $\Lambda$CDM model are shown for
$\beta_1=3.2$, for which QSOs alone are adequate for providing the UV
background at $z\ge4$.}
\label{fig:emiss}
\end{figure}

For any pair of $\epsilon_L$ and $r_0$, $\Gamma$ is specified through
eq.~(\ref{eq:Gammap}), and therefore $\bar\tau_\alpha$. Any pair that yields a
desired $\bar\tau_\alpha$, however, is not necessarily self-consistent. This is
because any given value of $\Gamma$ fixes not only $\bar\tau_\alpha$ but
$\langle\exp(-\tau_L)\rangle$, and therefore $r_0$, as well. It is thus
necessary to solve self-consistently for values of $r_0$ as a function of
$\epsilon_L$. We show the dependence of $r_0$ and $\bar\tau_\alpha$ on
$\epsilon_L$ for the results of our $\Lambda$CDM simulation in
Fig~\ref{fig:emiss_r0_tau}.
Here $\alpha_{\rm MG} \approx0$ has been taken to account for the hardening
of the radiation field; the associated uncertainty in the amount of hardening,
which will also depend on re-emission of Lyman continuum photons by the IGM
itself, introduces an uncertainty of $30-50\%$ in the estimates for
$\epsilon_L$. It is also assumed that the sources have turned on at very high
redshifts, so that the limit $z_{\rm on}\rightarrow\infty$ may be taken in the
expression for $l_{\rm H}$. For small values of $\epsilon_L$, the background
intensity is attenuation limited (case AL). For large values, the intensity is
limited by cosmological expansion (case CL). There is a rapid transition
between the two regimes. For the constant comoving density source case,
the transition exhibits critical behaviour, with a sudden transition from
case AL to case CL as the emissivity is increased:\ a sufficiently large
emissivity drives the attenuation length to such large values that attenuation
becomes negligibly important. As attenuation becomes less significant, the
mean \Lya optical depth rapidly declines as well. Although the transition is
less abrupt for the more gradual QSO evolution case, the ionization structure
of the IGM is still very sensitive to the source emissivity. This sensitivity
permits a tight bound to be set on the metagalactic emissivity.

We show in Fig~\ref{fig:emiss} the emissivity required to match the measured
values of $\bar\tau_\alpha=0.73$, 2.1 and $>5.1$ at the respective redshifts
$z\approx4$, 5 and 6. We use the results of Songaila \& Cowie (2002) for the
point at $z=4$, estimating the error from the spread in their two fits
to $\bar\tau_\alpha$. The measured values at $z\approx5$ and 6 are from
Becker \etal 2001, with their quoted errors, and only the 1$\sigma$ lower
limit is used at $z\approx6$ based on \Lya absorption. They set more
stringent limits on $\bar\tau_\alpha$ at $z\approx6$ using \Lyb and \Lyg
absorption, but these are somewhat less firm due to absorption modelling
uncertainties.

The estimate of $\epsilon_L$ for Lyman Break Galaxies at $z\approx3$
(Steidel \etal 2001) exceeds the required value at $z>4$ by nearly an order of
magnitude. If the proper emissivity were constant out to $z=4$, attenuation by
the IGM would become negligible, $\bar\tau_\alpha$ would be driven to very low
values and the IGM would become nearly transparent to \Lya photons. Either the
proper emissivity of LBGs declines dramatically between $z\approx3$ and 4,
or the emissivity of LBGs at $z\approx3$ has been overestimated. Ferguson,
Dickinson \& Papovich (2002) argue that indeed the star formation rate in
Lyman Break Galaxies declines sharply for $z>3$.

The estimated emissivities from QSOs assuming luminosity slopes at
the bright end of $\beta_1=2.58$, 3.2 and 3.41 are also shown. While QSOs
alone are not able to provide the required emissivity for $\beta_1=2.58$,
the preferred value from the Sloan (Fan \etal 2001a, 2001b), QSOs alone are
easily adequate for $\beta_1=3.41$, the value preferred by the 2dF at
lower redshift. We find that the predicted QSO emissivity is very sensitive to
$\beta_1$, scaling roughly as $\epsilon_L\sim\beta_1^{4.5}$. The QSO emissivity
becomes sufficient for $\beta_1>3.2$ for $z\ge4$ within the errors. The reason
for the sensitivity to $\beta_1$ is that as the luminosity function steepens,
the low luminosity QSOs are able to provide an increasing proportion of the
total emissivity. While the emissivity above $M^*_{1450}$ varies little with
$\beta_1$ due to the constraints imposed by the number of QSOs detected at
the bright end, the proportion of the total emissivity provided by QSOs
fainter than $M^*_{1450}$ increases from about 50\% to nearly 80\% as $\beta_1$
increases from 2.58 to 3.41. For $\beta_1=3.2$, 75\% of the total emissivity
is provided by QSOs fainter than $M^*_{1450}$. Thus most of the emissivity
is derived from QSOs that are not directly detected. Yet it would seem
unlikely that low luminosity QSOs, detected in lower redshift surveys, would
disappear altogether. Our estimates are based on the conservative assumption
that the faint QSOs continue to follow pure luminosity evolution, although
we have no direct evidence for this.

The attenuation length $r_0$ is predicted to decay exponentially with
redshift, $r_0\approx9\times10^4\exp\left[-1.5(1+z)\right]$~Mpc for
$4\le z\le6$ and $\beta_1=3.2$, as shown in the lower panel of
Fig~\ref{fig:emiss}. The mean number $\bar N_0$ of QSOs within the
attenuation volume similarly follows an exponential decay, $\bar
N_0\approx9\times10^{12}\exp\left[-4.35(1+z)\right]$, with fewer than
a single QSO on average per attenuation volume by $z=6$. If QSOs
dominate the UV background, then large fluctuations in the background
field and ionization level of the IGM will be introduced for $z>5$. We
next turn to the possible role played by a fluctuating background
field on the absorption statistics of the IGM.

%%%%%%%%%%%%%%%%%%%%%%%%%%%%%%%%%%%%%%%%%%%%%%%%%%%%%%%%%%%%%%%%%%%%%%%%%%%%%%%
\section{UV background fluctuations}
\label{sec:flucs}

The stochastic nature of the discrete sources of the UV metagalactic
background will introduce local fluctuations in the intensity of the
background. When these sources are sufficiently numerous, the
fluctuations will be small. This is expected to be the case for any
contribution from galaxies or other stellar sources (eg, intergalactic
star clusters), because of the large number density of these sources.
Because of their rarity, however, the contribution from
QSO sources will fluctuate at high redshifts. The degree of
fluctuation is determined by the number of sources within a
photoelectric attenuation volume, the region around the source to
which Lyman continuum photons may propagate on average before being
photoelectrically absorbed by the IGM.

In order to compute the probability distribution $f(J)$ of the UV
background, we begin by computing the distribution $f(J,\bar N)$ of
the UV background generated by an average of $\bar N$ sources within a
finite sphere of radius $R$.  Here $J$ designates the angle averaged
specific intensity of the radiation field at a given frequency and
point in space, which we take to be the origin of our coordinate
system. (We suppress the frequency subscript to simplify notation.) We
work in the Euclidean limit. For any given realisation of $N$ sources
with specific luminosities $L_k$ and distances $r_k$, $J$ is given by
\begin{equation}
J = \Sigma_{k=1}^N j_k,
\label{eq:Jsum}
\end{equation}
where $j_k=L_k\exp(-\tau_k)/ (4\pi r_k)^2$, $\tau_k=r_k/r_0$ is the optical
depth due to photoelectric absorption by the IGM, and $r_0$ is the
attenuation length.
The distribution of $J$ is given
straightforwardly by the method of characteristic functions (Kendall \&
Stuart 1969) when the contribution of each source is statistically
independent. The characteristic function of $f(J, N)$ is then (denoting
an average over all the sources by $\langle ...\rangle$),
\begin{eqnarray}
\hat f(t, N)&\equiv&\langle \exp(itJ)\rangle \\
&=& \int\, \exp(it\Sigma_{k=1}^N j_k)
\left[\Pi_{k=1}^N p(L_k, r_k)\, dL_k\, dr_k\right]. \nonumber
\label{eq:charJ}
\end{eqnarray}
Here $p(L_k, r_k)\, dL_k\, dr_k$ denotes the probability of source $k$
having a specific luminosity in the range $(L_k, L_k + dL_k)$ and
lying within a sphere of radius $R$ at a distance from the origin in
the range $(r_k, r_k+dr_k)$. It is given by
\begin{equation}
p(L_k, r_k)\, dL_k\, dr_k=\frac{4\pi r_k^2}{(4/3)\pi R^3}\frac{\Phi(L_k)}
{\int_{L_{\rm min}}^{L_{\rm max}}\, \Phi(L)\, dL}\, dL_k\, dr_k,
\label{eq:probjk}
\end{equation}
and is identical for all the sources. The expression then simplifies to
$\hat f(J, N)=I^N(t,R)$, where, defining $x=L/L^*$,
\begin{equation}
I(t,R) = \int_0^R\, dr \frac{3r^2}{R^3} \int_{L_{\rm min}/L^*}
^{L_{\rm max}/L^*}\, dx \phi(x) \exp(itj),
\label{eq:ItN}
\end{equation}
where we have used eq.~(\ref{eq:probjk}) for $p(L, r)$ (dropping the subscript
$k$), and $\phi(x)$ is the normalized luminosity function of the sources
\begin{equation}
\phi(x) = \frac{\Phi(xL^*) L^*}{\int_{L_{\rm min}}^{L_{\rm max}}\, \Phi(L) dL},
\label{eq:phiL}
\end{equation}
where $\Phi(L)$ is the luminosity distribution per unit luminosity
and unit (comoving) volume.

Inverting $\hat f(J,N)$ we obtain
\begin{equation}
f(J,N)=\frac{1}{2\pi}\int\, \exp(-itJ) I^N(t,R)\, dt.
\label{eq:probJN}
\end{equation}
Performing a Poisson average over $N$ sources with
mean $\bar N=(4\pi/3)R^3 \bar n$, where $\bar n = \int\, \Phi(L) dL$,
gives
\begin{eqnarray}
f(J,\bar N)&\equiv&\sum_{N=0}^{\infty}\frac{{\bar N}^N e^{-\bar N}}{N!}f(J,N)
\nonumber \\
&=&\frac{1}{2\pi}\int\, \exp(-itJ)
\exp\left\{\bar N\left[I(t,R)-1\right]\right\} dt.
\label{eq:probJ}
\end{eqnarray}

Zuo (1992a) obtained eq.~(\ref{eq:probJ}) using Markoff's method,
although the derivation is more straightforward as shown here. He then
presents solutions for the restricted case of an identical luminosity for
all sources, no absorption within the attenuation volume ($\tau_k=0$),
and $R=r_0$.

It is useful to recaste eq.~(\ref{eq:ItN}) using
the dimensionless variables $s=tJ^*$ and $\tau=r/r_0$,
where $J^*=L^*/ (4\pi r_0)^2$. The expression for $I(t,R)$ becomes
\begin{eqnarray}
\label{eq:IsN}
I(s,R) &=& 3\frac{r_0}{R}\int_0^{R/r_0}\, d\tau \tau^2\,
\int_{x_{\rm min}}^{x_{\rm max}}\, dx \phi(x)
\exp(isx\tau^{-2}e^{-\tau}) \nonumber \\
&=& \int_{x_{\rm min}}^{x_{\rm max}}\, dx \phi(x) \biggl\{\exp\left[isx
(r_0/R)^2e^{-R/r_0}\right] \\
\phantom{\biggl\{}& & \mbox{}+ isx\frac{\bar N_0}{\bar N}
\int_0^{R/r_0}\, d\tau
\exp\left(isx\tau^{-2}e^{-\tau}\right)e^{-\tau} (2+\tau)\biggr\}\nonumber,
\end{eqnarray}
where $x_{\rm min}=L_{\rm min}/L^*$,  $x_{\rm max}=L_{\rm max}/L^*$,
$\bar N_0=(4\pi/3)r_0^3\bar n$ is the average number of sources within
an attenuation volume, and we have performed an integration
by parts to obtain the final expression.

We may now take the limit $R\rightarrow\infty$ of eq.~(\ref{eq:IsN})
to obtain
\begin{equation}
\lim_{R\to\infty}\bar N\left[I(s,R)-1\right]=
is\bar N_0\int_{x_{\rm min}}^{x_{\rm max}}\, dx x\phi(x)G(sx),
\label{eq:Is}
\end{equation}
where we have defined the function
\begin{equation}
G(\omega)\equiv\int_{\xi=0}^\infty\, d\xi e^{i\omega\xi}\tau^3(\xi),
\label{eq:G}
\end{equation}
for which $\tau$ is related implicitly to $\xi$ through
$\xi=\tau^{-2}e^{-\tau}$. Using eqs.~(\ref{eq:probJ}) and (\ref{eq:Is}),
we obtain for the distribution function of $j=J/J^*$,
\begin{eqnarray}
f(j)&=&\lim_{R\to\infty}f(J,\bar N)J^* \nonumber \\
&=&\frac{1}{\pi}\int_0^\infty\, ds
\cos\left[s\bar N_0\int_{x_{\rm min}}^{x_{\rm max}}\, dx x\phi(x)
{\sl Re}~G(sx)-sj\right] \nonumber \\
& & \mbox{} \times\exp\left[-s\bar N_0
\int_{x_{\rm min}}^{x_{\rm max}}\, dx x\phi(x){\sl Im}~G(sx)\right].
\label{eq:probj}
\end{eqnarray}

The leading asymptotic behaviours of the real and imaginary parts
of $G(\omega)$ for $\omega\ll1$ are given by ${\sl
Re}~G(\omega)\sim\int_0^\infty\, d\xi\tau^3(\xi)=3$ and ${\sl
Im}~G(\omega)\sim\int_0^\infty\, d\xi\sin(\omega\xi)\xi^{-3/2}=
(2\pi\omega)^{1/2}$. These expressions may be used to derive the
leading asymptotic behaviour of $f(j)$ for $j\gg1$. We obtain
$f(j)\sim\frac{3}{4}\bar N_0\langle x^{3/2} \rangle j^{-5/2}$
$(j\gg1)$, found by rotating the integral over $s$ in
eq.~(\ref{eq:probj}) into the complex plane and using the method of
stationary phase. Here $\langle x^{3/2}\rangle$ is an average over the
luminosity function $\phi(x)$. The average value of $j$ is found
directly from eq.~(\ref{eq:probJ}) to be $\langle
j\rangle\equiv\int_0^\infty\, dJ (J/J^*) f(J,\bar N)=3\bar N_0\langle x\rangle
\left(1-e^{-R/r_0}\right)\rightarrow3\bar N_0\langle x\rangle$ for
$R\rightarrow\infty$. Using the asymptotic form for $f(j)$, we note that the
{\it rms} fluctuations of $j$ diverge at the bright end ($j_{\rm max}\gg1$)
like $j_{\rm max}^{1/4}$.

\begin{figure}
\begin{center}
\leavevmode \epsfxsize=3.3in \epsfbox{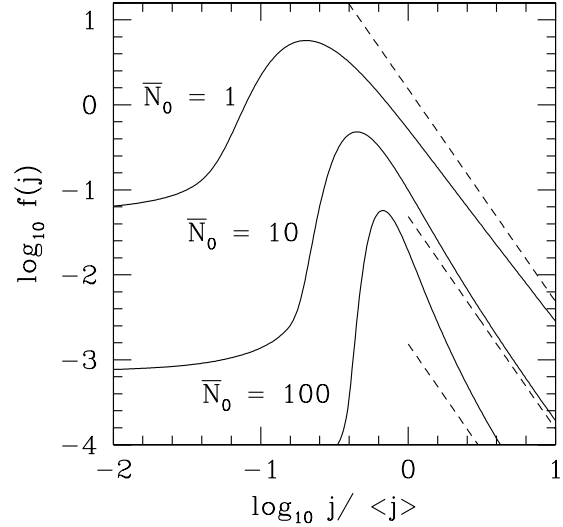}
\end{center}
\caption{The probability density $f(j)$ for the fluctuations in the
intensity $j$ for $\bar N_0=1$, 10 and 100 for the QSO luminosity function
with $\beta_1=2.58$ and $\beta_2=1.58$
(solid lines). Also shown are the corresponding asymptotic limits
$f(j)\sim \frac{3}{4}\bar N_0 \langle x^{3/2}\rangle j^{-5/2}$ (dashed
lines). The distributions peak at lower $j/\langle j\rangle$ for decreasing
$\bar N_0$.}
\label{fig:Jdist}
\end{figure}

In this paper we solve eq.~(\ref{eq:probj}) numerically using
the full QSO luminosity function as determined from observations. We show in
Fig~\ref{fig:Jdist} the resulting distributions for $\bar N_0=1$, 10 and 100,
using eq.~(\ref{eq:BLF}) with $\beta_1=2.58$ and $\beta_2=1.58$. The peak of
the distribution shifts toward lower intensity values as $\bar N_0$ decreases.

%%%%%%%%%%%%%%%%%%%%%%%%%%%%%%%%%%%%%%%%%%%%%%%%%%%%%%%%%%%%%%%%%%%%%%%%%%%%%%%
\section{The \Lya forest flux}
\label{sec:flux}

\subsection{The pixel flux distribution}
\label{sec:test_flux}

\begin{figure}
\begin{center}
\leavevmode \epsfxsize=3.3in \epsfbox{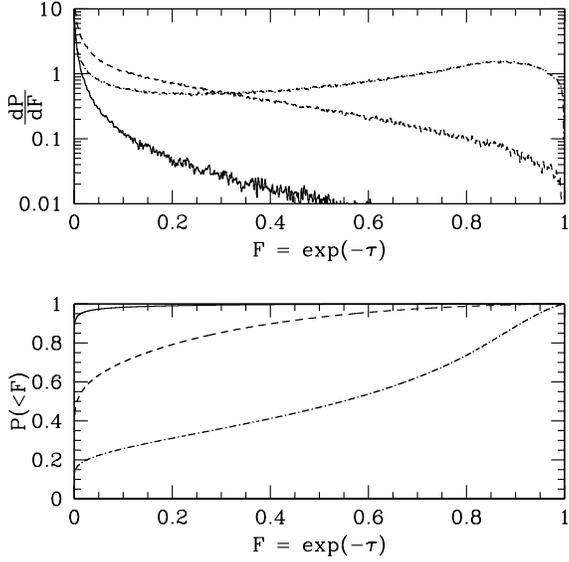}
\end{center}
\caption{Comparison of flux per pixel distributions (top panel) and cumulative
flux distributions (bottom panel)
at $z=4$ (dot-dashed), 5 (dashed) and 6 (solid), normalized
to the measured values of $\bar\tau_\alpha$. The distributions sharply steepen
at vanishing flux levels at high redshift.}
\label{fig:fdist}
\end{figure}

The distribution function of flux per pixel has proven to be an extremely
sensitive discriminator between \Lya forest models. The number $N$ of pixels
available in a high resolution spectrum, like those from the Keck HIRES or
VLT UVES, is on the
order of $10^4$, so that the cumulative distribution of pixel fluxes may
be measured to an {\it rms} accuracy of $\sim N^{-0.5}\approx0.01$, although
the finite width of the absorption features tends to increase the {\it rms}
somewhat because of the resulting statistical dependence between pixels
(Meiksin \etal 2001). We show the pixel flux distributions from the simulations
at $z=4,$ 5 and 6 in Fig~\ref{fig:fdist}. Because of the rapidly rising values
of $\bar\tau_\alpha$ with redshift, the distributions become increasingly
peaked at low flux values.

\begin{figure}
\begin{center}
\leavevmode \epsfxsize=3.3in \epsfbox{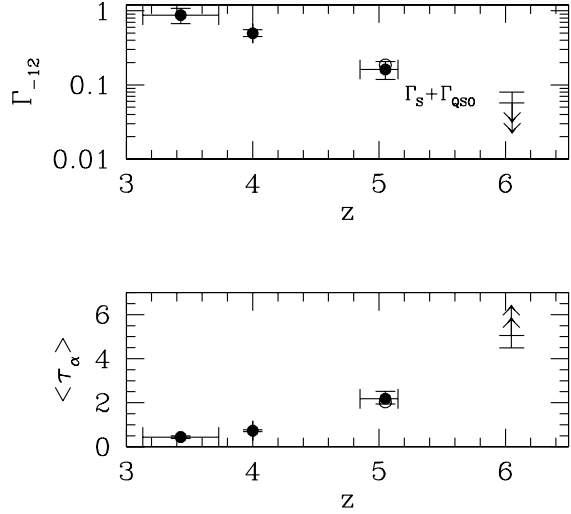}
\end{center}
\caption{(Top panel)\ Estimates for the required mean \HI
photoionization rate $\bar\Gamma_{-12}$, in units of $10^{-12}\, {\rm
s^{-1}}$, required to match the measured $\bar\tau_\alpha$ at
different redshifts. The lowest redshift point is from Meiksin \etal
(2001). The solid points are for an assumed homogeneous
photoionization background. The open circle at $z\approx5$ takes into
account fluctuations in the background contributed by QSO sources for
$\beta_1=3.2$, also including a small contribution from a smoothly
distributed component. At $z\approx6$, the lower upper limit is for
the case of a homogeneous photoionization background. The limit marked
$\Gamma_{\rm S}+ \Gamma_{\rm QSO}$ includes both a smooth component
and the effects of QSO fluctuations for $\beta_1=3.2$. (Bottom
panel)\ The measured $\bar\tau_\alpha$ at different redshifts. The
lower points at $z=5$ and 6 show the values of $\bar\tau_\alpha$ that
would result assuming the same mean photoionization rate required by
QSOs to match the measured values of
$\bar\tau_\alpha$, but neglecting the effect of the background
fluctuations. The fluctuations substantially boost $\bar\tau_\alpha$
at $z\approx6$.}
\label{fig:Gamma}
\end{figure}

\begin{figure}
\begin{center}
\leavevmode \epsfxsize=3.3in \epsfbox{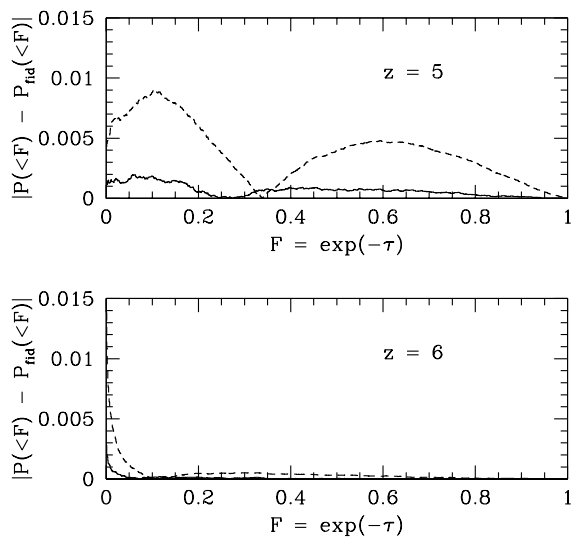}
\end{center}
\caption{Differences of the cumulative distributions of flux per pixel for the
$\Lambda$CDM simulation with different assumptions for the origin of the
photoionization background, at $z=5$ (top panel) and $z=6$ (bottom panel).
The flux distributions are all normalized to the measured values of
$\bar\tau\alpha$. The cumulative distribution differences are with reference to
the fiducial flux per pixel distribution for which a uniform photoionization
background is assumed. The solid lines show the difference when fluctuations
in a background dominated by QSOs are taken into account, adopting
$\beta_1=3.2$. The dashed lines show the
difference assuming $T_0=10^4\kel$ instead of $T_0=2\times10^4\kel$ for
the temperature-density relation, and assuming a homogeneous UV background.
}
\label{fig:cumd}
\end{figure}

Fluctuations in the UV background will affect the pixel flux distribution in
two ways:\ a shift in its mean and a tilt in its shape. Matching the measured
mean flux to the value predicted for a given set of cosmological parameters
fixes the mean $\Gamma$ for a simulation, so that neglecting fluctuations may
lead to a mis-estimate of the mean $\Gamma$.

At $z=4$, the number of QSOs in an attenuation volume is sufficiently large
that the effects of background fluctuations will be negligible. We therefore
only  consider the higher redshift cases. We concentrate on the case
$\beta_1=3.2$ for the bright end of the QSO luminosity function, since the
estimates in Table~\ref{tab:LF} suggests QSOs alone are able to meet the
emissivity requirements to match the measured values of $\bar\tau_\alpha$.

We show in Fig~\ref{fig:Gamma} (upper panel) the required ionization
rates, both with and without the effects of a fluctuating UV
background. At $z=5$, the fluctuations have only a small effect on the
required $\Gamma$, increasing it slightly above the case of smoothly
distributed sources, for which $\Gamma_{-12}=0.16$.  Once the effect
of background fluctuations is included, it is no longer possible for
QSOs alone to maintain the ionization of the IGM for $\beta_1=3.2$;
unless the metagalactic UV spectrum hardens to $\alpha_{\rm MG}<0$, a
small additional smooth component of $\Gamma_{\rm S, -12}=0.025$ must
be added. The smooth component is an increase by 15\% over the mean
QSO ionization rate, a reasonable amount for the diffuse contribution
expected to arise from the re-radiation of Lyman continuum photons by
the IGM itself. The difference is larger at $z\approx6$.  The total
ionization rate must now increase from a mean of
$\bar\Gamma_{-12}=0.057$, to reproduce $\bar\tau_\alpha=5.1$, to
0.080, of which $\sim40$\% is made up of a smoothly distributed
component, again reasonable for re-radiation by the IGM. In the lower
panel of Fig~\ref{fig:Gamma}, a comparison is shown between the
measured values of the mean \Lya optical depth and the values that
would be obtained assuming the same required mean QSO rates as in the
upper panel, but neglecting the role of the background
fluctuations. The significant boost of the estimated mean \Lya optical
depth by the UV background fluctuations demonstrates the need for
taking into account the role of the fluctuations at high redshifts
($z\gsim5$). In particular, this shows that care must be taken in
ascribing an abrupt rise in the mean \Lya optical depth to the
reionization epoch, as done by Becker \etal (2001), as such a rise
may also be due in part to the UV background fluctuations
resulting from a small number of sources in an attenuation volume.

We note that the ionization rates in Fig~\ref{fig:Gamma} decay exponentially
with redshift from $z\gsim3.5$, paralleling the decline in the number counts
of QSO sources beyond this redshift.

A comparison of the cumulative distributions of flux per pixel
is shown in Fig~\ref{fig:cumd}. We show the differences in the cumulative
distributions, as would be used in the Kolmogorov-Smirnov (KS) test.
The fiducial models for $z=5$ and $z=6$ assume a homogeneous UV background.
We compare the homogeneous case with the case of $\beta_1=3.2$, taking into
account both the fluctuations in the UV background from the QSOs and the small
additional smooth component as found above. The effect on the cumulative
distributions is small and beyond the current detection
limits of the Keck and VLT, except possibly for
large sample sizes. The differences, however, may be measurable with an ELT
class telescope. We also compare the cumulative distributions (for smoothly
distributed sources) for two assumptions for the IGM temperature. For the
fiducial case, $T_0=2\times10^4$~K in eq.~(\ref{eq:eos}) was assumed.
Decreasing
$T_0$ to $10^4$~K significantly alters the cumulative distribution at $z=5$,
although has a smaller effect at $z=6$. In principle, the uncertainty in the
IGM temperature, which corresponds to a change in line-width of the absorption
features, may be corrected for by adjusting the line-widths to match the
measured Doppler parameter distribution or wavelet coefficient distribution
(Meiksin 2000; Meiksin \etal 2001), so that this uncertainty may not be a
limiting factor in detecting the effect of the fluctuations on the pixel flux
distribution. We note that lowering the IGM temperature increases the radiative
recombination rate and so increases the required ionization rate from
$\Gamma_{-12}=0.16$ to 0.22 at $z=5$ and from $\Gamma_{-12}=0.057$ to 0.069
at $z=6$.

\subsection{Correlations in the UV photoionization background}
\label{sec:xi_J}

\begin{figure}
\begin{center}
\leavevmode \epsfxsize=3.3in \epsfbox{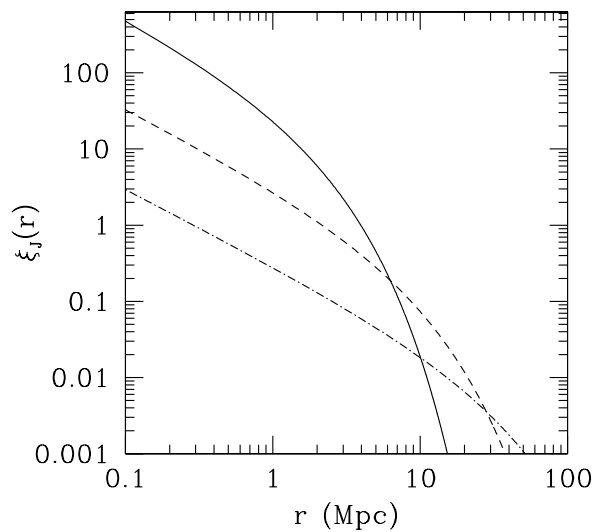}
\end{center}
\caption{The correlation function of the photoionization UV background
generated by QSO sources alone assuming $\beta_1=3.2$, at $z=4$ (dot-dashed),
5 (dashed) and 6 (solid).
}
\label{fig:xi_J}
\end{figure}

Zuo (1992b) showed that the discreteness of the sources of the UV background
in the presence of attenuation
will result in correlations between the background intensity at points
separated in space. His result for the correlation function may be
expressed as
\begin{equation}
\xi_J(r)=\frac{1}{3\bar N_0}\frac{\langle x^2\rangle}{\langle x\rangle^2}
\frac{r_0}{r}\int_{r/r_0}^\infty\, du\, \frac{1}{u}\log\frac{u+r/r_0}{u-r/r_0}
e^{-u},
\label{eq:xi_J}
\end{equation}
where the averages $\langle\dots\rangle$ are performed over the
dimensionless QSO luminosity function $\phi(x)$ (eq.~[\ref{eq:phiL}]).
In Fig~\ref{fig:xi_J}, we show the correlation function for
$\beta_1=3.2$, for which nearly the entire UV background may be
provided by QSOs. The correlation lengths at $z=4$, 5 and 6 are 0.29,
2.2 and 4.0 Mpc (proper), respectively, corresponding to the velocity
differences (in the QSO restframe) of 125, 1250 and 2850\kms. The
background correlations will induce correlations in the pixel fluxes
as well, and so serve as a possible means for detecting the
discreteness of the sources. The correlations will become somewhat
diluted by any contribution from a more smoothly distributed
component, as may arise from the diffuse background re-radiated by the
IGM. We treat the effect of UV background correlations in a separate
paper (Meiksin \& White 2003).

\subsection{Reionization of the IGM}
\label{sec:reion}

In this paper, we have considered QSOs as the possible sources that dominate
the UV background at $z\lsim6$. This does not require the QSOs to have
initiated the ionization itself, however. The ratio of the Hubble time
$t_{\rm H}$ to the time for full radiative recombination in our $\Lambda$CDM
model is
\begin{equation}
t_{\rm H}n_{\rm H}\alpha_A\approx0.024\left(\frac{\Omega_bh^2}{0.020}\right)
(1+z)^{3/2}\left(\frac{\rho}{\bar\rho}\right)^{0.65},
\label{eq:tHtrec}
\end{equation}
where we have used eq.~(\ref{eq:eos}) and approximated the radiative
recombination coefficient as $\alpha_A\approx4\times10^{-13}
(T/10^4\,{\rm K})^{-0.7}\,{\rm cm^3\, s^{-1}}.$ At $z=6$, the ratio exceeds
unity only for $\rho/\bar\rho\gsim3.5$. At the mean IGM density,
the ratio is less than unity for $z<11$. If an early generation
of AGN or stars pre-ionized the IGM, for instance, it would be possible for
the generation of QSOs at $z<6$ to maintain the ionization before the IGM
had time to fully recombine. It is none the less interesting to
address the question of whether these QSOs would have been sufficient to
have done the reionization as well. The
number density of ionizing photons generated by the QSOs over a Hubble time,
assuming $\beta_1=3.2$ and $\eta=1/4$, is
$n_\gamma(z)=6.8\times10^{-4}(1+z)^{-5/4}$\cm3. Comparing this with the
total number density of hydrogen atoms $n_{\rm H}(z)$ gives
\begin{equation}
\frac{n_\gamma(z)}{n_{\rm H}(z)}\approx4000\left(\frac{\Omega_bh^2}{0.020}
\right)^{-1}(1+z)^{-17/4}.
\label{eq:ngnH}
\end{equation}
This crosses unity at $z=6.04$. If these QSOs were responsible for
the reionization of the IGM, then they only just could have done
so at the highest redshifts to which they have been seen. In the
presence of clumping, reionization by the QSOs becomes problematic.
If, however, the IGM is still partially ionized at $z\approx6$
following an earlier epoch of reionization, then the QSOs would be
adequate for renewing the ionization of the IGM, as the required
number of ionizing photons scales like the neutral fraction of
hydrogen atoms.

%%%%%%%%%%%%%%%%%%%%%%%%%%%%%%%%%%%%%%%%%%%%%%%%%%%%%%%%%%%%%%%%%%%%%%%%%%%%%%%
\section{Summary}
\label{sec:summary}

We have investigated the dependence of the ionization structure of the
IGM at high redshifts ($z\gsim4$) on the emissivity of the sources of
the UV photoionization background in the context of a $\Lambda$CDM
universe. We show that the ionization structure is highly sensitive to
the emissivity of the sources at the Lyman continuum edge. For low
emissivity values, the intensity of the UV background is limited by
the attenuation of Lyman continuum photons by the IGM. For large values of
the emissivity, the IGM becomes essentially transparent to Lyman continuum
photons, and the UV background is limited instead by cosmological
expansion and the age of the sources. We show there is a rapid
transition between these two extremes with increasing emissivity.
If the UV background is dominated by Lyman Break Galaxies at $z\gsim4$,
then their proper emissivity must dramatically decline from the value
estimated at $z\approx3$ (Steidel \etal 2001), or they would render the
IGM nearly transparent at the Lyman edge and result in a much smaller
mean \Lya optical depth than measured.

We have also explored the possibility that the UV background is
dominated by QSO sources at high redshift. We show that within the
observational errors on the high redshift QSO luminosity function,
QSOs may provide a substantial component to the UV background, and may
even dominate it. If they do dominate, then their sparseness will
result in substantial fluctuations in the UV background. We derive the
distribution function for the fluctuating intensity in an infinite
universe including the effects of attenuation, and show that the
distribution significantly broadens for $z\gsim5$. The fluctuations
in the UV background will increase the estimated \Lya optical depth
over the value that would be estimated assuming a homogeneous
background. This increases the demand placed on the ionization rate
required to reproduce a measured mean \Lya optical depth and should be
taken into account when comparing with estimates of the QSO
emissivity. It also shows that care must be taken in attributing a
sudden rise in the \Lya optical depth to the reionization epoch
(Becker \etal 2001), since a sudden rise may result instead from
a rapidly decreasing attenuation length and the resulting UV background
fluctuations associated with the reduced
number of sources that dominate the background. There is a secondary
effect of the fluctuations on the pixel flux distribution of the \Lya
forest, resulting in a small distortion. Although the effect is too
small to be measured in a single QSO spectrum from existing facilities
like the Keck or VLT, it may be detectable by combining many spectra,
or may need to wait until the arrival of a future generation of
extremely large telescopes.

The large mean \Lya optical depth measured by the Sloan at $z\approx6$
(Becker \etal 2001) requires fewer than a single QSO within an
attenuation volume on average. This is a significant measurement in
that it implies that if QSOs dominate the UV background, there will be
large fluctuations in the mean \Lya optical depth from region to
region, over spatial scales of $\sim4$~Mpc, the expected correlation
length of the background radiation field produced by the QSOs at
$z\approx6$. This offers a possible means of distinguishing between
models for the origin of the UV background dominated by either rare
sources like QSOs, or by relatively more copious sources like galaxies.
We pursue this topic in a companion paper (Meiksin \& White 2003).

We have also shown that if QSOs dominate the UV background at $z\approx6$,
they would just have had sufficient time to provide an adequate number
of Lyman continuum photons to ionize the hydrogen, but not at larger
redshifts. The estimate, however, is based on the minimal counting
requirement of the production of one ionizing photon per hydrogen
atom, and neglects the possible role of clumping of the gas. On the
other hand, simulations show that much of the volume of the IGM will
be underdense and so more readily photoionized. If the IGM were
ionized at an earlier phase and had not yet fully recombined by
$z\approx6$, then even in the presence of clumping, QSOs may be
adequate for renewing the ionization of the IGM. We also note that the
UV background fluctuations we expect, if QSOs dominate the ionization
of hydrogen, will apply to helium ionization as well, for which it is
even more certain that QSOs were the sources of reionization. A
complete description of the reionization of the IGM will require
accounting for the nature of the ionizing sources, their immediate
environments, and the inhomogeneous structure of the IGM. Reviews of
recent work in this evolving subject are provided by Madau (2000) and
Loeb \& Barkana (2001).

In order to match the statistics of observed spectra in detail, it
will ultimately be necessary to incorporate radiative transfer into
the simulations both to allow for the expected spatial correlations in
the UV ionizing background as well as to reproduce the correct gas
temperatures (and line widths), particularly if helium is completely
photoionized at only moderate redshifts. Several other physical
effects may also be important, such as heating sources in addition to
QSOs, or galactic feedback like winds.

%%%%%%%%%%%%%%%%%%%%%%%%%%%%%%%%%
\bigskip
\section*{acknowledgments}

The simulations used here were performed on the IBM-SP2 at the National
Energy Research Scientific Computing Center.
M.W. was supported by the NSF, NASA and a Sloan Foundation Fellowship.
 
%%%%%%%%%%%%%%%%%%%%%%%%%%%%%%%%%

%%%%%%%%%%%%%%%%%%%%%% TABLES %%%%%%%%%%%%%%%%%%%%%
\begin{table}
\centerline{\begin{tabular}{|c|c|c|c|c|c|c|} \hline
$z$&& $M^*_L$ &&& $\epsilon_L$ \\
$\beta_1=$ & 2.58 & 3.20 & 3.41 & 2.58 & 3.20 & 3.41\\
\hline 
\hline
4 & -22.86 & -23.98 & -24.23 & 3.23 & 7.96 & 9.77 \\ \hline
5 & -22.11 & -23.44 & -23.73 & 2.68 & 8.19 & 10.54 \\ \hline
6 & -21.52 & -23.02 & -23.35 & 2.36 & 8.62 & 11.52 \\ \hline
\end{tabular}}
\caption{Break luminosity $M^*_L$ at the Lyman edge and proper emissivity
at the Lyman edge required to match the Sloan QSO counts at different
redshifts, assuming a QSO luminosity function slope at the bright end of
$\beta_1=2.58$, 3.20 and 3.41 and a homogeneous UV background.
The units of the emissivity are $10^{26}\emunits$.
}
\label{tab:LF}
\end{table}


\begin{thebibliography}{}

%\bibitem[1986]{BBKS86}
%        Bardeen J.~M., Bond J.~R., Kaiser N., Szalay A.~S., 1986,
%        \apj, 304, 15

\bibitem[]{beck01}
	Becker R.~H., \etal, 2001, \aj, 122, 2850

\bibitem[]{bon97}
        Bond J.~R., Wadsley J.~W., 1997, in Petitjean P., Charlot S., eds,
        Structure and Evolution of the Intergalactic Medium from QSO
        Absorption Line Systems. Editions Fronti\`eres, Paris, p. 143

\bibitem[]{boy88}
        Boyle B.~J., Shanks T., Peterson B.~A., 1988, \mnras, 235, 935

\bibitem[]{boy00}
        Boyle B.~J., Shanks T., Croom S.~M., Smith R.~J., Miller L.,
	Loaring N., Heymans C., 2000, MN, 317, 1014

\bibitem[]{bruz01}
	Bruzual G., 2001, \apsss, 277, 221

%\bibitem[]{bry98}
%        Bryan G.~L., Machacek M., Anninos P., Norman M.~L., 1999, \apj,
%        517, 13

%\bibitem[]{bun97}
%        Bunn E.~F., White M., 1997, \apj, 480, 6

\bibitem[]{cen94}
        Cen R., Miralda-Escud\'e J., Ostriker J.~P., Rauch M.,
        1994, \apj, 437, L9

\bibitem[]{cro02}
	Croft R.~A.~C., Hernquist L., Springer V., Westover M.,
	White M., 2002, preprint (astro-ph/0204460)

%\bibitem[1998]{cwkh98}
%	Croft R.~A.~C., Weinberg D.~H., Katz N., Hernquist L., 1998, \apj,
%	495, 44

%\bibitem[]{dav97}
%        Dav\'e R., Hernquist L., Weinberg D.~H., Katz N., 1997, \apj,
%        477, 21

\bibitem[]{fan01a}
        Fan X., \etal, 2001a, \apj, 121, 54

\bibitem[]{fan01b}
        Fan X., \etal, 2001b, \apj, 122, 2833

\bibitem[]{fan02a}
        Fan X., \etal, 2002, \apj, 123, 1247

\bibitem[]{fan02b}
        Fan X., \etal, 2003, preprint (astro-ph/0301135)

\bibitem[]{fdp02}
	Ferguson H.~C., Dickinson M., Papovich C., 2002, \apj, 569, 65

\bibitem[]{fioc97}
	Fioc M., Rocca-Volmerange B., 1997, A\&A, 326, 950

%\bibitem[]{gh98}
%        Gnedin N.~Y., Hui L., 1998, MN, 296, 44

\bibitem[]{HM96}
	Haardt F., Madau P., 1996, \apj, 461, 20

\bibitem[]{her96}
        Hernquist L., Katz N., Weinberg D., Miralda-Escud\'e J.,
        1996, \apj, 457, L51

\bibitem[1988]{HocEas}
	Hockney R.~W., Eastwood J.~W., 1988, Computer Simulation Using
        Particles, Adam Hilger, Bristol

%\bibitem[]{HG97}
%        Hui L., Gnedin N.~Y., 1997, \mnras, 292, 27

\bibitem[]{KS69}
	Kendall M.~G., Stuart A., 1969, The Advanced Theory of Statistics,
	Vol. 1. Charles Griffin \& Co., London

\bibitem[]{LB01}
	Loeb A., Barkana R., 2001, \araa, 39, 19

%\bibitem[]{m00}
%        Machacek M.~E., Bryan G.~L., Meiksin A., Anninos P., Thayer D.,
%        Norman M.~L., Zhang Y., 2000, \apj, 532, 118~(M00)

\bibitem[]{mad00}
	Madau P., 2000, RSPTA, 358, 2021

\bibitem[]{mei94}
        Meiksin A., 1994, ApJ, 431, 109

\bibitem[]{mei00}
        Meiksin A., 2000, \mnras, 314, 566

\bibitem[]{mm93}
	Meiksin A., Madau P., 1993, \apj, 412, 34

%\bibitem[1999]{MWP}
%	Meiksin A., White M., Peacock J.~A., 1999, MNRAS, 304, 851

\bibitem[2001]{MW01}
	Meiksin A., White M., 2001, \mnras, 324, 141

\bibitem[2002]{MW02}
	Meiksin A., White M., 2003, in preparation

\bibitem[]{mbm00}
        Meiksin A., Bryan G.~L., Machacek M.~E., 2001, \mnras, 327, 296

\bibitem[]{nett02}
	Netterfield C.~B., \etal, 2002, \apj, 571, 604

\bibitem[]{omea01}
	O'Meara J.~M., Tytler D., Kirkman D., Suzuki N., Prochaska J.~X.,
	Lubin D., Wolfe A.~M., 2001, \apj, 552, 718

\bibitem[]{pdf01}
	Papovich C., Dickinson M., Ferguson H.~C., 2001, \apj, 559, 620

\bibitem[]{perc01}
	Percival W.~J., \etal, 2001, \mnras, 327, 1297

%\bibitem[1995]{pmk85}
%	Petitjean P., M\"ucket J.~P., Kates R.~E., 1995, A\&A, 295, L9

\bibitem[]{pry02}
	Pryke C., Halverson N.~W., Leitch E.~M., Kovac, J., Carlstrom J.~E.,
	Holzapfel W.~L., Dragovan M., 2002, \apj, 568, 46

\bibitem[]{sch00}
	Schaye J., Theuns T., Rauch M., Efstathiou G., Sargent W.~L.~W.,
	2000, \mnras, 318, 817

\bibitem[]{sc02}
	Songaila A., Cowie L.~L., 2002, \aj, 123, 2183

\bibitem[]{spa01}
	Steidel C.~C., Pettini M., Adelberger K.~L., 2001, \apj, 546, 665

\bibitem[]{stom01}
	Stompor R., \etal, 2001, \apj, 561, L7

\bibitem[]{sza01}
	Szalay A., \etal, 2001, preprint (astro-ph/0107419)

%\bibitem[]{the98}
%        Theuns T., Leonard A., Efstathiou G., 1998, \mnras, 297, L49

\bibitem[]{the98a}
        Theuns T., Leonard A., Efstathiou G., 1998a, \mnras, 297, L49

\bibitem[]{theu98b}
        Theuns T., Leonard A., Efstathiou G., Pearce F.~R., Thomas P.~A.,
        1998b, \mnras, 301, 478

%\bibitem[1999]{whi99}
%	White M., 1999, \mnras, 310, 511

\bibitem[]{zha95}
        Zhang Y., Anninos P., Norman M.~L., 1995, \apj, 453, L57

\bibitem[]{zman97}
        Zhang Y., Anninos P., Norman M.~L., Meiksin A.,
        1997, \apj, 485, 496

%\bibitem[]{zman98}
%        Zhang Y., Meiksin A., Anninos P., Norman M.~L.,
%        1998, \apj, 495, 63

\bibitem[]{zhe97}
	Zheng W., Kriss G.~A., Telfer R.~C., Grimes J.~P.,
	Davidsen A.~F., 1997, \apj, 475, 469

\bibitem[]{zuo92a}
	Zuo L., 1992a, \mnras, 258, 36

\bibitem[]{zuo92b}
	Zuo L., 1992b, \mnras, 258, 45

\bibitem[]{zp93}
	Zuo L., Phinney, E.~S., 1993, \apj, 418, 28

\end{thebibliography}
\end{document}